\documentclass{ws-ijmpa}

\usepackage[super]{cite}
\usepackage{xcolor}
\usepackage[verbose,hypertexnames=false]{hyperref}
\hypersetup{colorlinks=false,allbordercolors=blue,pdfborderstyle={/S/U/W 1}}
\begin{document}

\markboth{Z.~A.~Abdelrazik, M. Csanád}{Lévy-stable pion emission sources in Au+Au collisions at $\sqrt{s_{\rm NN}}=200$ GeV from the vHLLE+SMASH hybrid model}

%%%%%%%%%%%%%%%%%%%%% Publisher's Area please ignore %%%%%%%%%%%%%%%
%
\catchline{}{}{}{}{}
%
%%%%%%%%%%%%%%%%%%%%%%%%%%%%%%%%%%%%%%%%%%%%%%%%%%%%%%%%%%%%%%%%%%%%

\title{Lévy-stable pion emission sources in Au+Au collisions at $\sqrt{s_{\rm NN}}=200$ GeV from the vHLLE+SMASH hybrid model}

\author{Zeinab A. Abdelrazik$^{1,2}$, Máté Csanád$^{3}$}

\address{$^{1}$Department of Physics, Faculty of Science, Cairo University\\
Giza, 12613, Egypt\\
zabdelrazik@cu.edu.eg}

\address{$^{2}$Center for High Energy Physics(CHEP-FU), Fayoum University,\\ El-Fayoum 63514, Egypt}

\address{$^{3}$Department of Atomic Physics, ELTE Eötvös Loránd University,\\ Budapest, 1117, Hungary\\csanad@elte.hu}

\maketitle

\begin{history}
\received{Day Month Year}
\revised{Day Month Year}
\accepted{Day Month Year}
\published{Day Month Year}
\end{history}

\begin{abstract}
Revealing the space-time evolution of the quark–gluon plasma created in heavy-ion collisions is a central challenge, as this information is encoded only in the properties of the particles emitted at freeze-out. In this work, we reconstruct the pion emission source in Au+Au collisions at $\sqrt{s_{\rm NN}}=200$ GeV (0–5\% centrality) simulated with the vHLLE+SMASH hybrid framework, combining a Glauber Monte Carlo initial state, viscous hydrodynamic expansion $(3+1)$D, and a hadronic afterburner that describes resonance decays and hadronic rescattering. The Hanbury Brown--Twiss (HBT) interferometry is used to characterise the pair source for identical-pion pairs, while in the case of simulations it can be extracted directly from the last point of scattering of each final particle. The projected pair-separation distributions in the Bertsch--Pratt out, side, and long directions are analyzed with a three-dimensional L\'evy-stable parameterization. Over the investigated fit range, which extends to approximately $100$~fm, a single L\'evy-stable function provides a satisfactory description of the distributions, with the extracted Lévy index $\alpha \approx 1.5-1.6$, below the Gaussian value of 2, slowly increasing with $m_T$ . At larger pair separations, the total source departs from the fitted distribution, consistent with contributions from long-lived resonance decays forming the halo. Short-lived resonance decays and hadronic rescattering can contribute to the non-Gaussian shape of the distribution via the Lévy-walk phenomenon. In addition, the three extracted HBT radii $R_{\rm out}$, $R_{\rm side}$, and $R_{\rm long}$ decrease with $m_T$, in qualitative agreement with the expected signatures of collective expansion. These results demonstrate that L\'evy-stable femtoscopy provides an integrative framework that constructs a useful link between the hadronic-stage dynamics of the hybrid model and experimentally accessible femtoscopic observables.
\end{abstract}


\ccode{PACS numbers: 25.75.Gz, 25.75.$-$q, 12.38.Mh}

\section{Introduction}
\label{sec:intro}

Heavy-ion collisions are an effective experimental tool for understanding
nuclear forces and the hadronic matter structure; they also reveal the
properties of the deconfined state of strongly interacting matter, the
quark-gluon plasma (QGP), that is believed to have filled the early universe
microseconds after the Big Bang. The nuclear matter goes through extreme conditions, with temperatures about $10^5$ times higher than in the core of the Sun and energy densities several times that of normal nuclear matter \cite{ref1}; these conditions lead to the formation of the deconfined QGP, followed by hadronization into the hadrons observed in the detectors. The evolution proceeds through several tightly
coupled stages: initial-state geometry fluctuations, hydrodynamic expansion
of the QGP, hadronization, and a subsequent hadronic rescattering stage that
affects the experimental observables at the freeze-out stage. The QGP therefore cannot be probed directly: accessing the early stages requires combining theoretical modeling with experimental analysis, since their signatures may be obscured by the subsequent hadronic evolution.

There are several theoretical approaches to modeling initial-state
conditions and treating final-state observables of heavy-ion collisions to
interpret the properties of the QGP matter. For instance, a relativistic hydrodynamics model treats the QGP as an expanding fluid characterized by transport coefficients such as the shear viscosity to entropy density ratio $\eta/s$, while microscopic transport models describe the late-stage hadronic evolution
through resonance formation and particle rescattering in the dilute hadronic
gas phase. Reconstructing the space-time evolution of the collision system
directly from the particles measured in detectors remains one of the central challenges of the field. Therefore, the
research focuses on using femtoscopy (HBT interferometry), which provides access to the space-time structure of the particle-emitting source via quantum-statistical correlations.

In this work, we study identical-pion femtoscopic correlations in
events generated with the vHLLE--SMASH hybrid transport model, which couples
a viscous relativistic hydrodynamic description of the early, strongly
coupled QGP phase to a hadronic afterburner capable of resolving individual
rescattering and resonance-decay processes after partialization. We analyze the pair-separation distributions and fit them with a Lévy-stable parametrization, aiming to identify the microscopic origins of
non-Gaussian emission and to quantify how the extracted stability index and
source radii evolve with transverse mass, providing a bridge between the
underlying transport dynamics and experimental femtoscopic observables.

\section{Theoretical Framework}
\label{sec:theory}

\subsection{Femtoscopic interferometry}

Hanbury Brown--Twiss (HBT) interferometry, pioneered in high-energy physics by Goldhaber and collaborators~\cite{ref4} is a quantum-mechanical approach
based on the momentum correlations between identical particles, used to
probe the spatial and temporal structure of the emission source. The
two-particle correlation function is defined as the ratio between the
probability of jointly observing two particles with a given relative
momentum and the probability expected for independent emission, as follows:
\begin{equation}
C_2(\mathbf{q},\mathbf{K}) = \frac{P(\mathbf{p}_1,\mathbf{p}_2)}{P(\mathbf{p}_1)\,P(\mathbf{p}_2)}\, ,
\label{eq:c2def}
\end{equation}
where
\begin{equation}
\mathbf{q} = \mathbf{p}_1 - \mathbf{p}_2\, , \qquad
\mathbf{K} = \frac{1}{2}(\mathbf{p}_1 + \mathbf{p}_2)\, ,
\label{eq:qK}
\end{equation}
with $\mathbf{q}$ the relative momentum and $\mathbf{K}$ the average pair
momentum. The correlation function is related to the pair-separation distribution $D(\mathbf{r},\mathbf{K})$ through the Koonin–Pratt equation\cite{ref5,Koonin:1977fh}:
\begin{equation}
C_2(\mathbf{q},\mathbf{K}) = \int S(\mathbf{r},\mathbf{K})\, |\Psi_{\mathbf{q}}(\mathbf{r})|^2\, d^3r
\label{eq:koonin-pratt}
\end{equation}

where $\Psi_{\mathbf{q}}(\mathbf{r})$ is the relative
two-particle wave function with $\mathbf{r} = \mathbf{r}_1 - \mathbf{r}_2$ being the pair separation. For non-interacting identical bosons (such as
pions), $|\Psi_{\mathbf{q}}(\mathbf{r})|^2$ is replaced by
$\cos(\mathbf{q}\cdot\mathbf{r})$, leading to the Fourier transform expression:
\begin{equation}
C_2(\mathbf{q},\mathbf{K}) \simeq 1 + \int D(\mathbf{r},\mathbf{K})\, \cos(\mathbf{q}\cdot\mathbf{r})\, d^3r
\label{eq:c2fourier}
\end{equation}

Therefore, for identical bosons such as pions, quantum interference enhances
the correlation function at low relative momentum, so that the width of the
measured correlation function encodes the size and shape of the emitting
source at freeze-out.

\subsection{LCMS frame and coordinate decomposition}\label{ssec:lcms}

Tracking the two-particle correlations in HBT interferometry requires a
reference frame and coordinate system to resolve the spatial source
geometry. First, the pair trajectories are analyzed in the longitudinally
comoving system (LCMS). For each pair, the LCMS is obtained by a boost
along the beam direction such that the sum of the longitudinal momenta
vanishes,
\begin{equation}
p_{1z} + p_{2z} = 0\, 
\label{eq:lcms}
\end{equation}
in order to isolate the intrinsic longitudinal structure of the source
without the influence of the overall longitudinal motion of the pair.
Simultaneously, to resolve the three-dimensional shape of the emitting
source inside the LCMS frame, both the relative momentum and the pair separation are decomposed into the Bertsch--Pratt\cite{Bertsch:1988db,Pratt:1986cc} components, and the pair separation components in LCMS are written as ~\cite{Kincses:2024lnv}
\begin{equation}
\rho = \left(\rho_{\rm out}, \rho_{\rm side}, \rho_{\rm long}\right)
\label{eq:rho}
\end{equation}
where the out direction is chosen parallel to the pair transverse momentum,
the side direction is perpendicular to it in the transverse plane, and the
long direction lies along the beam axis. The normalized distribution \(D(\rho,K_T)\) then describes the separation of identical-pion pairs satisfying the kinematic cuts in bins of $K_T$, which is expressed via the pair transverse mass $m_T=\sqrt{K_T^2 + m^2}$ with $m$ being the mass of the individual (and identical) particles; furthermore, the direct notion of $K_T$ dependence is omitted below.

\subsection{L\'evy-stable source parametrization}

Several effects can lead to a non-Gaussian, power-law-tailed source, such as anomalous diffusion in the expanding hadron gas, resonance decays, event averaging, or critical phenomena;~\cite{levy-phenix,Csanad:2024hva} in such cases the standard Gaussian-diffusion
picture may not fully capture the underlying dynamics, and the source may be described by Lévy-stable distributions. This is manifested as superdiffusive transport motion, described by L\'evy
walks~\cite{Kincses:2024lnv}, which involve occasional very long
propagation steps that cannot be produced by ordinary Brownian diffusion. This results in an elliptically contoured Lévy-stable pair separation distribution~\cite{Csorgo:1999sj,Kincses:2025izu}:
\begin{equation}
\mathcal{L}(\alpha,R^2,{\bf r})=\frac{1}{(2\pi)^3} \int d^3{\bf q} e^{i{\bf q} {\bf r}} e^{-\frac{1}{2}|{\bf q}^T R^2 {\bf q}|^{\alpha}},
\label{eq:levysource}
\end{equation}
where $\alpha$ is the Lévy index of stability, $R^2$ is the matrix of source radii (HBT radii, or sometimes also called homogeneity lengths), while ${\bf q}$ is an integration variable. In the Bertsch--Pratt decomposition $R^2$ is often assumed to be diagonal, with the square root of the diagonal components being $R_{\rm out}$, $R_{\rm side}$, $R_{\rm long}$. The value $\alpha = 2$ reproduces the familiar Gaussian source, while $\alpha < 2$ signals non-Gaussian, anomalous
transport: smaller values of $\alpha$ correspond to heavier tails. The correlation function becomes ~\cite{Csorgo:1999sj}
\begin{equation}
C_2(\mathbf{q}) = 1 + \lambda \cdot
\exp\!\left(
-\left|\sum_{i \in \{\rm out,side,long\}} (q_i R_i)^2\right|^{\alpha/2}
\right) 
\label{eq:levycorrfunc}
\end{equation}
where the additional intercept parameter $\lambda$ appears\cite{Csorgo:1994in}, which represents the weight of the source in the small-distance range of up to 100 fm~\cite{Kincses:2025izu}. This parametrization allows the geometric size of the source and the
degree of anomalous transport to be quantified simultaneously. In this work, we employ Eq.~\eqref{eq:levysource} as a parameterization for the modeled $D(\rho)$ distributions. 
The projections of $D(\boldsymbol{\rho})$ are obtained from a single
three-dimensional elliptically contoured Lévy-stable pair-source
distribution.\\

The one-dimensional out, side, and
long projections of the $D(\boldsymbol{\rho})$ distribution have
been fitted simultaneously, with a common stability index $\alpha$, while their
direction-dependent Lévy scales are $R_{\rm out}$, $R_{\rm side}$, and
$R_{\rm long}$. If the single-pion source has Lévy scale
$R_{\rm single}$, the separation of two independently emitted pions has scale
$R_{\rm pair}=2^{1/\alpha}R_{\rm single}$. In this work, the reported $R_{\rm out}$,
$R_{\rm side}$, and $R_{\rm long}$ that are plotted in Figure~\ref{fig:mtdep} are the pair-source scales fitted
directly to $D(\boldsymbol{\rho})$, as defined in Eq.~(7), rather
than scales of the single-pion source.

\subsection{Modeling framework}

The collision system under study is Au+Au collisions at a center-of-mass
energy per nucleon of 200~GeV, simulated using the vHLLE--SMASH hybrid transport
framework~\cite{vhlle,smash,hybrid}, with 1000 events generated for the present analysis. The hybrid framework connects the
macroscopic fluid dynamics with microscopic hadronic transport within three
modeling stages, beginning with the determination of the initial collision geometry
and energy profile based on the Glauber model~\cite{glissando}.
The generated initial state is then evolved hydrodynamically with vHLLE, a
$(3+1)$-dimensional viscous relativistic hydrodynamic code~\cite{vhlle}, to
simulate the space--time evolution prior to hadronization. The evolution
uses a lattice-QCD equation of state with $N_f = 3$
(\texttt{Laine\_nf3.dat}, \texttt{eosType}~$=1$), matched at
$T_\mathrm{ch} = 149$~MeV to a hadron-resonance-gas equation of state
(\texttt{eosTypeHadron}~$=1$), with baryon and strangeness susceptibilities
taken from lattice QCD (\texttt{chiB.txt}, \texttt{chiS.txt}) and a constant
shear-viscosity-to-entropy ratio $\eta/s = 0.15$.

Once the medium becomes sufficiently dilute, particlization is performed on a hypersurface of constant energy density using the hadron sampler, version 3.0, and the subsequent hadronic stage, including resonance formation, rescattering and decays, is simulated with the SMASH hadronic transport setup, version 3.3. The final particle coordinates and momenta produced by this hybrid framework are then used to construct identical-pion pairs and to calculate femtoscopic observables.

\begin{figure}[t]
\centerline{\includegraphics[width=\textwidth]{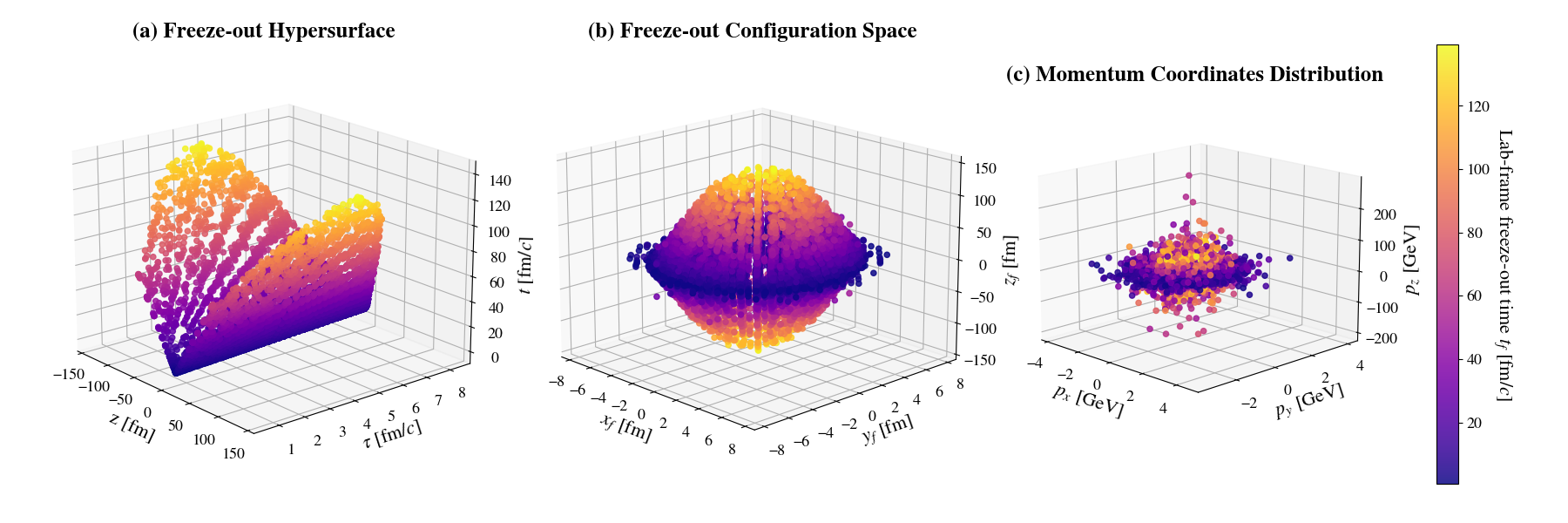}}
\caption{
Particle ensemble sampled from the hydrodynamic freeze-out hypersurface through
Au+Au collisions at $\sqrt{s_{NN}}=200$~GeV and $0$--$5\%$ centrality, within
the vHLLE--SMASH hybrid framework. Panel~(a) shows the freeze-out surface
traced by the sampled emission points in Milne coordinates,
$\tau=\sqrt{t^{2}-z^{2}}$. Panels~(b) and~(c) show a subsample of particles
in coordinate space and momentum space, respectively, coloured according
to their freeze-out time $t_f$ in the laboratory frame.}
\label{fig:vhlle}
\end{figure}

\section{Results and Discussion}
\label{sec:results}
The full freeze-out hypersurface of the Au+Au collision has been extracted
using the vHLLE--SMASH hybrid framework, and the space-time and momentum
coordinate configuration for a subsample of particles is visualized in
Figure~\ref{fig:vhlle}.
All identical-pion pairs are subjected to the same kinematic cuts,
$0.15 < p_T < 1\ \mathrm{GeV}/c$ and $|\eta| < 1$, and are binned
according to pair transverse momentum $K_T$. In addition, only pairs with small relative momentum in the longitudinally co-moving system (LCMS) are kept,
$Q_{\mathrm{LCMS}}=\sqrt{q_{\mathrm{out}}^{2}+q_{\mathrm{side}}^{2}+q_{\mathrm{long}}^{2}} <  \sqrt{0.2~\mathrm{GeV}\; m_T}$, where the $m_T=\sqrt{K_T^{2}+m_\pi^{2}}$ evaluated at the the transverse momentum of the pair $K_T$ bin.

For each pair, the two
particles' space-time coordinates are taken at their individual emission vertices. The difference
$(\Delta t, \Delta x, \Delta y, \Delta z)$ is transformed to the LCMS
and decomposed in the Bertsch--Pratt coordinates as described in Sec.~\ref{ssec:lcms}.\\
In
particular, the out and long components depend explicitly on the time
difference~\cite{Kincses:2024lnv}:
\begin{align}
\rho_{\mathrm{out}}
&=
\Delta x\cos\phi+\Delta y\sin\phi
-
\frac{K_T}{K_0^2-K_z^2}
\left(K_0\Delta t-K_z\Delta z\right), \\
\rho_{\mathrm{side}}
&=
-\Delta x\sin\phi+\Delta y\cos\phi, \\
\rho_{\mathrm{long}}
&=
\frac{K_0\Delta z-K_z\Delta t}
{\sqrt{K_0^2-K_z^2}}.
\end{align}
\\

\begin{figure}[t]
\centerline{\includegraphics[width=12.8cm]{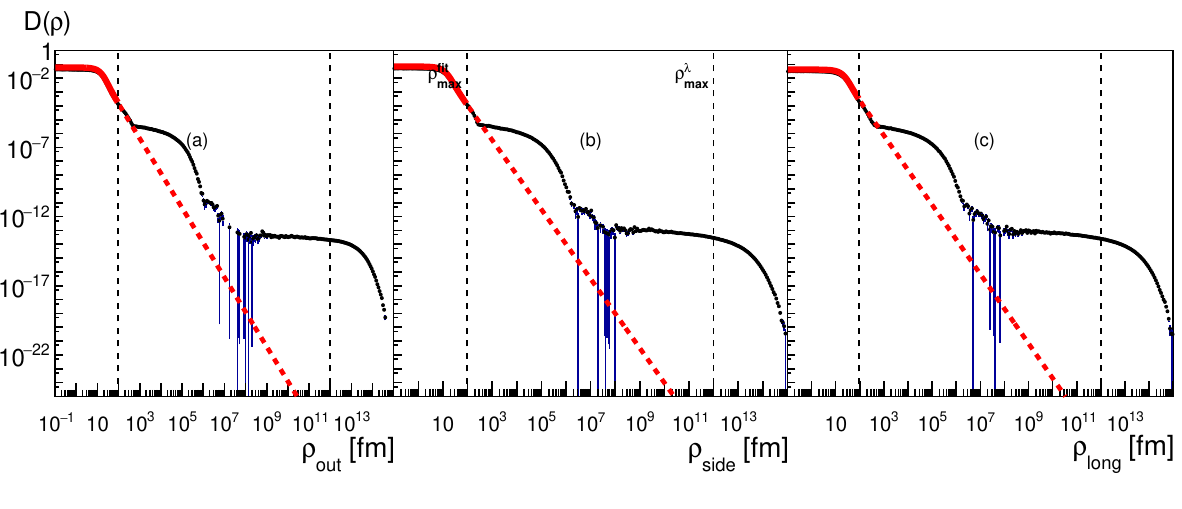}}
\caption{
Pair separation distributions $D(\rho)$ for identical pion pairs in the
$K_T=0.22$--$0.27$~GeV/$c$ interval, shown for the three spatial components
$\rho_{\mathrm{out}}$, $\rho_{\mathrm{side}}$, and $\rho_{\mathrm{long}}$.
The solid red lines represent the L\'evy-stable fit function, while the dashed lines indicate the extrapolations of the
fitted functions beyond the fitted range. The vertical lines around $10^9$ fm are caused by the very small number of hits in these bins of the histogram and the corresponding Poisson uncertainties.}
\label{fig:drho}
\end{figure}

Thus, $\rho_{\mathrm{out}}$ is not simply the transverse spatial
separation: its time-difference term can affect both the magnitude and
the tail of the out projection. The pair-source distributions $D(\boldsymbol{\rho})$ and their one-dimensional out-side-long projections are constructed for several $K_T$ intervals; an example is shown in Figure~\ref{fig:drho}. The pair-separation cut is applied jointly to the three components, through the modulus $|\rho| = \sqrt{(\rho_{out}² + \rho_{side}² + \rho_{long}²)} < \rho_{max}^{\lambda}$,
where $\rho_{\max}^{\lambda}$ is the cutoff that defines the correlation parameter $\lambda$. Based on experimental analyses of daughter tracks from weak decays of long-lived hadrons
($K^0_S$, $\Lambda$, $\Sigma$, $\Xi$, $\Omega$, with $c\tau$ of a few cm) can be rejected
by selections on the distance of closest approach (DCA) to the primary vertex. The accepted amount of
this feed-down depends on the experiment's track and DCA cuts. Therefore, the cut-off $\rho_{\max}^{\lambda}=1~\mathrm{mm}$ is a normalization choice for the source:
only pairs with $|\rho|<\rho_{\max}^{\lambda}$ enter the normalization of $D(\rho)$, so
the pairs from long-lived decays at larger separations do not contribute to $\lambda$.
It is intended to mimic the finite track resolution and DCA acceptance of the
measurement, but it is not equivalent to applying a $1~\mathrm{mm}$ DCA cut to each
daughter track. This distinction matters when the extracted $\lambda$ is compared to data,
because the simulated and experimental samples must have compatible feed-down
acceptances. For this reason, $\lambda$ is also evaluated for the larger cutoffs of
$5~\mathrm{mm}$, $5~\mathrm{cm}$ and $50~\mathrm{cm}$.\\

The distributions for three components exhibit heavy tails extending to very large
separations, originating from the decays of long-lived hadrons and from the long
free-streaming trajectories in the hadronic stage. Since SMASH may not decay the weakly decaying hadrons
% ($K^0/\bar K^0$, $\Lambda$, $\Sigma$, $\Xi$, $\Omega$) or the $\eta$ at the end of the
run, these are preprocessing decayed with the branching ratios and lifetimes of the
particle tables; the decay point is sampled from $\beta\gamma c\tau$ of the parent. The three projected pair-separation distributions are fitted simultaneously
with the appropriate projections of the 3D L\'evy-stable pair source from Eq.~\eqref{eq:levysource}, which are themselves 1D L\'evy-stable distributions with the same $\alpha$ and the respective radii. The fit is performed in a limited interval,
$0.1~\mathrm{fm}\le\rho\le\rho_{\max}^{\mathrm{fit}}(m_T)$, whose upper limit depends on
the transverse mass by factor $\sqrt{{2500}/{m_T}}$, evaluated at the mean transverse mass $\langle m_T\rangle$ of the pairs in each $K_T$ bin. The maximum fit range decreases from  100 fm with increasing the $\langle m_T\rangle$ bin, therefore the fit range follows the shrinking source size
and stays roughly an order of magnitude above the fitted radii. Beyond
$\rho_{\max}^{\mathrm{fit}}$ the fitted function is only extrapolated (dashed lines in
Fig.~\ref{fig:drho}). Within this region, the fit
provides an effective description of the source without requiring a separate background from short-lived resonances; the adequacy of a single fit function
over the fit region should not be confused with the appearance of a core--halo structure in the complete source at much larger distances. The non-Gaussian
shape within the fitted region reflects the combined effects of short-lived
resonance decays and hadronic re-scattering, while at larger  distances the total
pair source departs from the fitted L\'evy distribution, consistent with the
contribution of long-lived resonance decays to the far halo.\\

A further feature of the reconstructed source is a sparsely populated region of the $D(\rho)$ distribution at intermediate separations. This structure reflects the discrete lifetime spectrum of the parent resonances: the core region is dominated by short-lived parents, pions from very long-lived parents populate the far halo, and the decay-length interval between intermediate parents, such as the $\eta$, and very long-lived parents, such as the $K^0_S$ and $\Lambda$, is weakly populated. The long-distance halo is further under-populated by the incomplete evolution of the decay chains of long-lived parents, such as the $\eta'$, whose daughters include the longer-lived $\eta$, $\rho^0$, and $\omega$. Because the missing pairs would have fallen at separations above the cutoff, this deficiency can bias the correlation-strength parameter $\lambda$ upward. Consequently, the quoted $\lambda$, obtained with the normalisation cutoff $\rho_{\max}^{\lambda}$, should be interpreted with the caveat that this cutoff corresponds to the finite source reconstruction range, as the correlation length $\lambda$ depends on the completeness of the reconstructed resonance contributions. The fitted L\'evy shapes and HBT radii are expected to be less sensitive to such missing contributions, because they are primarily constrained by the core region. Overall, the L\'evy-stable parameterization captures the shape of the pair-separation distributions in all three directions, with residual deviations that may reflect an incomplete treatment of certain resonance contributions in the hadronic afterburner.

\begin{figure}[t]
\centerline{\includegraphics[width=12.8cm]{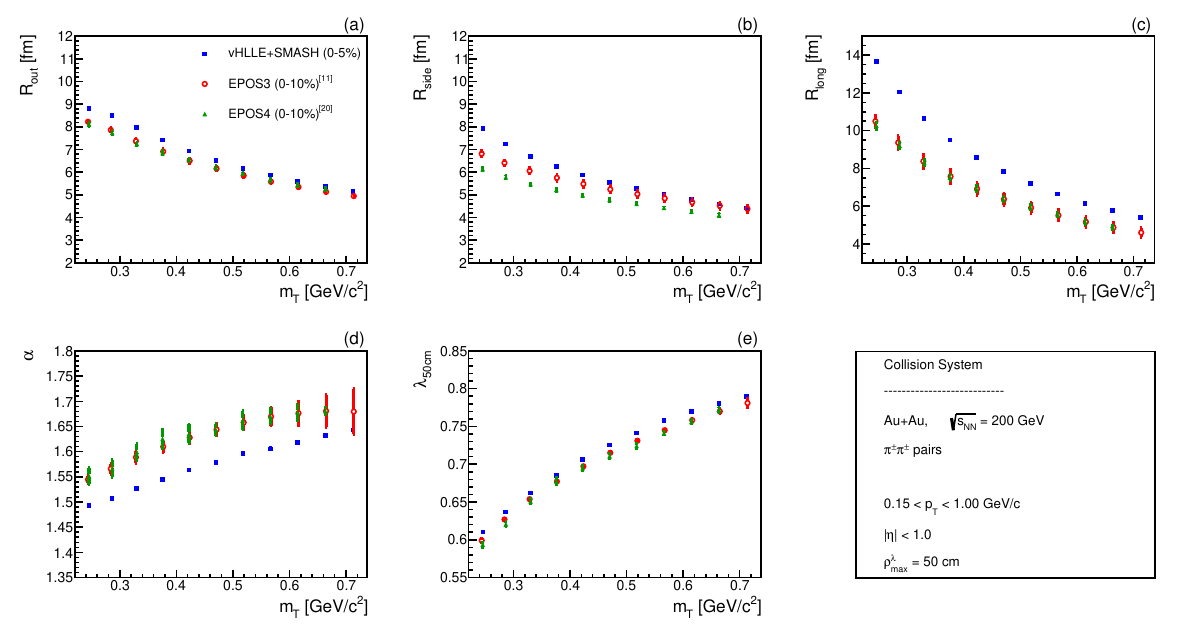}}
\caption{ Transverse-mass ($m_\mathrm{T}$) dependence of the L\'evy fit
  parameters for identical pion pairs in Au+Au collisions at
  $\sqrt{s_\mathrm{NN}}=200$~GeV. The vHLLE+SMASH results (0--5\% centrality) are compared
  with that of EPOS3~\cite{Kincses:2025izu} and EPOS4~\cite{huang2026epos4}  at 0--10\% centrality.
  Panels (a)--(c) show the three L\'evy radii $R_\mathrm{out}$,
  $R_\mathrm{side}$ and $R_\mathrm{long}$, panel (d) shows the L\'evy
  stability index $\alpha$, and panel (e) shows the correlation strength
  parameter $\lambda$, evaluated at $\rho_\mathrm{max}^{\lambda}=50$~cm.
  The points are plotted at the average $m_\mathrm{T}$ of the pairs in each
  $K_\mathrm{T}$ bin.
}
\label{fig:mtdep}
\end{figure}

It is necessary to examine the dependence of the L\'evy index $\alpha$, the correlation
strength $\lambda$, and the HBT radii on the transverse mass $m_T$, shown in
Fig.~\ref{fig:mtdep}, where the vHLLE+SMASH results for Au+Au collisions at 0--5\%
centrality are compared with EPOS3 and EPOS4 at 0--10\% centrality. The vHLLE+SMASH events
are sampled from the freeze-out hypersurface of a single hydrodynamic evolution; their
uncertainties therefore include only the sampling and hadronic-afterburner fluctuations. While , EPOS3 and EPOS4 are
event-by-event simulations with fluctuating initial conditions over a wider centrality
interval (0--10\%). \\

The HBT radii are shown in panels (a)--(c), in which all three radii decrease gradually with increasing
$m_T$ that driven by hydrodynamic-like flow and the associated space-momentum correlations. In
vHLLE+SMASH, $R_{\rm out}$ falls from $8.8$ to $5.2$~fm, $R_{\rm side}$ from $7.9$ to
$4.4$~fm and $R_{\rm long}$ from $13.7$ to $5.4$~fm across the average $m_T$ range.
The longitudinal radius shows the strongest decrease, reflecting its particular
sensitivity to the longitudinal expansion and to the freeze-out dynamics in the
longitudinal direction. The agreement with the results from simulated events from the EPOS3 and EPOS4 at 0-10 ${\%}$ centrality degree.
For $R_{\rm out}$ the three simulated frameworks are close: vHLLE+SMASH is larger by about
$0.7$~fm  at low $m_T$, and the difference shrinks to about $0.2$~fm at the
highest $m_T$. For $R_{\rm side}$ vHLLE+SMASH is larger by about $1.1$~fm than EPOS3 and
$1.8$~fm than EPOS4 at low $m_T$, and for $m_T\gtrsim0.5~\mathrm{GeV}/c^{2}$ it agrees
with EPOS3, while EPOS4 stays lower by about $0.3$~fm. The largest difference is in
$R_{\rm long}$, where vHLLE+SMASH exceeds both EPOS results by about $3$~fm at low $m_T$ and still by about $0.8$~fm at the highest $m_T$. While the EPOS4 yields systematically smaller $R_{\rm side}$ and $R_{\rm long}$ at low $m_T$ than EPOS3, consistent with the model-generation changes in the treatment of the longitudinal dynamics and of the hadronic stage; since these two ingredients are precisely what set $R_{\rm long}$ and part of $R_{\rm side}$, the ordering of the three curves encodes how differently the frameworks distribute the emission over time and along the beam direction.\\

The stability index $\alpha$ (panel~d) increases gradually with $m_T$, from
$\alpha\simeq1.49$ at $\langle m_T\rangle\simeq0.25~\mathrm{GeV}/c^{2}$ to
$\alpha\simeq1.64$ at $0.71~\mathrm{GeV}/c^{2}$. This trend is consistent with
higher-$m_T$ pions being emitted earlier from the hotter core regions under the strong
influence of collective expansion, so that they experience fewer long random propagation
paths. In all three directions $\alpha<2$, so the heavy tails are present globally and
are not confined to a single direction. EPOS3 and EPOS4 show the same trend but at
systematically higher values, with $\alpha\simeq1.55$ rising to $\simeq1.68$. The panel~(e) shows, the correlation strength $\lambda$ due to the cutoff
$\rho_{\max}^{\lambda}=50$~cm. It increases monotonically from about $0.61$ to $0.79$ and
agrees with both EPOS3 and EPOS4 over the whole $m_T$ range. This increase reflects the
core--halo structure of the source: at higher $m_T$ a larger fraction of pion pairs
originates from the hydrodynamic core, while the relative contribution of long-lived
decays decreases. Because the radii are extracted from the core region within the chosen
fit range, the long-lived halo affects the intercept parameter $\lambda$ without
appreciably modifying the radii. Since $\lambda$ depends on the normalization cutoff
$\rho_{\max}^{\lambda}$, the same cutoff of 50~cm is used for all three frameworks. However, the offset between the frameworks is attributed to the single freeze-out hypersurface used in vHLLE+SMASH and to the different centrality classes.

\label{sec:conclusion}\section{Conclusion}

The space-time structure of the particle-emitting source formed in Au+Au
collisions was simulated within the vHLLE--SMASH hybrid framework and studied
with femtoscopic HBT interferometry of identical pion pairs. 

The reconstructed pair-separation distributions exhibit pronounced, non-Gaussian tails that are described by a single Lévy-stable source function, which effectively fits the separation distribution range up to approximately $100$~fm, without eliminating the physical core–halo structure of the source. The contribution of long-lived resonances appears as a departure from the fitted region at larger separations, while the short-lived resonances and hadronic rescattering shaped the Lévy source region.

The extracted HBT radii indicate that the vHLLE--SMASH hybrid framework is qualitatively consistent with the collective
expansion behavior expected in heavy-ion collisions, with the longitudinal
radius $R_{\rm long}$ showing the strongest $m_T$ dependence, reflecting its
particular sensitivity to the longitudinal expansion of the system.
Furthermore, the research will extent to address the residual resonance-decay channels remaining
after the hadronic afterburner, in order to improve the modeling of the
complete core-halo structure of the source. In addition, higher-statistics
simulations are planned for a more precise extraction of the L\'evy
parameters, together with a direct quantitative comparison with experimental femtoscopic observables
\cite{levy-star,levy-cms,levy-na61}.

\section*{Acknowledgments}

We sincerely thank the organizing committee of the 18\textsuperscript{th} Workshop on Particle Correlations and Femtoscopy (WPCF 2026), held at ELTE, Budapest, for the opportunity to present this work. This research was partly supported by the NKFIH grant NKKP 152097.

\section*{ORCID}
\noindent Zeinab Abdelhalim Abdelrazik -- \url{https://orcid.org/0009-0000-1691-4622}
\noindent Mate Csanad -- \url{https://orcid.org/0000-0002-3154-6925}


\begin{thebibliography}{0}

\bibitem{ref1}
R.~Pasechnik and M.~\v{S}umbera, {\it Universe} {\bf 3}, 7 (2017),
\url{https://doi.org/10.3390/universe3010007}.

\bibitem{ref4}
G.~Goldhaber, S.~Goldhaber, W.~Y.~Lee and A.~Pais, {\it Phys. Rev.} {\bf 120}, 300--312 (1960),
\url{https://doi.org/10.1103/PhysRev.120.300}.

\bibitem{ref5}
S.~Pratt, {\it Phys. Rev. Lett.} {\bf 53}, 1219--1221 (1984),
\url{https://doi.org/10.1103/PhysRevLett.53.1219}.

\bibitem{Koonin:1977fh}
S.~E.~Koonin, {\it Phys. Lett. B} {\bf 70}, 43-47 (1977)
\url{https://doi.org/10.1016/0370-2693(77)90340-9}.

\bibitem{Bertsch:1988db}
G.~Bertsch, M.~Gong and M.~Tohyama,
%``Pion Interferometry in Ultrarelativistic Heavy Ion Collisions,''
{\it Phys. Rev. C} \textbf{37}, 1896-1900 (1988),
\url{https://doi.org/10.1103/PhysRevC.37.1896}.

\bibitem{Pratt:1986cc}
S.~Pratt,
%``Pion Interferometry of Quark-Gluon Plasma,''
{\it Phys. Rev. D} \textbf{33}, 1314-1327 (1986),
\url{https://doi.org/10.1103/PhysRevD.33.1314}.

\bibitem{Kincses:2024lnv}
D.~Kincses, M.~Nagy and M.~Csan{\'a}d,
%``L{\'e}vy walk of pions in heavy-ion collisions,''
{\it Commun. Phys.} \textbf{8}, no.1, 55 (2025),
\url{https://doi.org/10.1038/s42005-025-01973-x}.

\bibitem{Csanad:2024hva}
M.~Csan{\'a}d and D.~Kincses,
%``Femtoscopy with L{\'e}vy Sources from SPS through RHIC to LHC,''
{\it Universe} \textbf{10}, no.2, 54 (2024),
\url{https://doi.org/10.3390/universe10020054}.

\bibitem{levy-phenix}
A.~Adare {\it et al.} (PHENIX Collab.), {\it Phys. Rev. C} {\bf 97}, 064911 (2018),
\url{https://doi.org/10.1103/PhysRevC.97.064911}.

\bibitem{Csorgo:1999sj}
T.~Csörg\H{o},
%``Particle interferometry from 40-MeV to 40-TeV,''
{\it APH N.S., Heavy Ion Physics} {\bf 15}, 1–80 (2002),
\url{https://doi.org/10.1007/978-94-011-4126-0_8}.

\bibitem{Kincses:2025izu}
D.~Kincses, E.~Árpási, L.~Kovács, M.~Nagy and M.~Csanád,
%``Three-dimensional sizes and shapes of pion emission in heavy-ion collisions,''
{\it Eur. Phys. J. C} \textbf{86}, no.4, 333 (2026),
\url{https://doi.org/10.1140/epjc/s10052-026-15526-1}

\bibitem{Csorgo:1994in}
T.~Csörg\H{o}, B.~Lörstad and J.~Zimányi,
%``Bose-Einstein correlations for systems with large halo,''
{\it Z. Phys. C} \textbf{71}, 491-497 (1996),
\url{https://doi.org/10.1007/s002880050195}

\bibitem{vhlle}
I.~Karpenko, P.~Huovinen and M.~Bleicher, {\it Comput. Phys. Commun.} {\bf 185}, 3016
(2014), \url{https://doi.org/10.1016/j.cpc.2014.07.010}.

\bibitem{smash}
J.~Weil {\it et al.} (SMASH Collab.), {\it Phys. Rev. C} {\bf 94}, 054905 (2016),
\url{https://doi.org/10.1103/PhysRevC.94.054905}.

\bibitem{hybrid}
J.~Mohs, M.~Ege, H.~Elfner and M.~Mayer, {\it Phys. Rev. C} {\bf 105}, 034906 (2022),
\url{https://doi.org/10.1103/PhysRevC.105.034906}.

\bibitem{glissando}
W.~Broniowski, M.~Rybczy\'nski and P.~Bo\.zek, {\it Comput. Phys. Commun.} {\bf 180},
69 (2009), \url{https://doi.org/10.1016/j.cpc.2008.07.016}.

\bibitem{levy-star}
D.~Kincses {\it et al.} (STAR Collab.), {\it Universe} {\bf 10}, 102 (2024),
\url{https://doi.org/10.3390/universe10020102}.

\bibitem{levy-cms}
A.~Tumasyan {\it et al.} (CMS Collab.), {\it Phys. Rev. C} {\bf 109}, 024914 (2024),
\url{https://doi.org/10.1103/PhysRevC.109.024914}.

\bibitem{levy-na61}
H.~Adhikary {\it et al.} (NA61/SHINE Collab.), {\it Eur. Phys. J. C} {\bf 83}, 919
(2023), \url{https://doi.org/10.1140/epjc/s10052-023-11111-x}.

\bibitem{huang2026epos4}
Y. Huang, M. Molnar, D. Kincses, and M. Csanad,
``Excitation function of femtoscopic Lévy source parameters of pion pairs in EPOS4,''
{\it Phys. Lett. B} {\bf 876}, 140423 (2026),
\url{https://doi.org/10.1016/j.physletb.2026.140423}.

\end{thebibliography}
\end{document}